\documentclass[12pt]{article}
\usepackage{amsmath}
\usepackage{amssymb}
\usepackage{amsfonts}
\usepackage{nicefrac}
\usepackage[colorinlistoftodos,prependcaption, textsize=small]{todonotes}
\usepackage{setspace}
\usepackage[T1]{fontenc}
\usepackage{times}
\usepackage{palatino} 
\usepackage{lmodern}
\usepackage[latin1]{inputenc}
\usepackage{epsfig}
\usepackage[english]{babel}
\usepackage{color}
\usepackage{graphicx}
\usepackage{dcolumn}
\usepackage{moreverb}
\usepackage{cite}
\usepackage{bbm}

\usepackage[all,cmtip]{xy}

\usepackage[extra]{tipa}

\usepackage{stackengine}
\stackMath
\newcommand\tsup[2][2]{%
 \def\useanchorwidth{T}%
  \ifnum#1>1%
    \stackon[-1.3ex]{\tsup[\numexpr#1-1\relax]{#2}}{\mathchar"307E}%
  \else%
    \stackon[-1ex]{#2}{\mathchar"307E}%
  \fi%
}

\numberwithin{equation}{section}

\title{Cubic algebras, induced representations and general solution of the exceptional Laguerre equation $X_1$}
\author{Ian Marquette\\ 
{\small  Department of Mathematical and Physical Sciences, La Trobe University,} \\ 
{\small Bendigo 3552, Victoria Australia} 
}

\date{ }
\begin{document}
\baselineskip=22pt plus 1pt minus 1pt
\maketitle

\begin{abstract} 
We consider the case of exceptional Laguerre polynomials $X_1$ of type I, II and III, their ordinary differential equations and the problem of finding general solution beside the polynomial part. We will develop an algebraic approach based on the Schrodinger form of the problem and associate representations of the underlying spectrum generating algebra. We use the Darboux-Crum transformation to construct ladder operators of fourth order for the case of the exceptional Laguerre polynomials $X_1$ of type I, II and III. We then obtain all zero modes for the lowering and raising operators. We construct the induced representation for the linearly independent solutions, including the polynomial states. Those states forming the general solution are important non only in the construction of wider set of physical states satisfying different boundary conditions, but also used in context of getting isospectral deformations as they allow often to overcome obstruction as several Wronskian constructions of Hamiltonian lead to only formal Darboux transformations. Our approach, allows to provide a completely algebraic construction of the two linearly independent solutions of the ordinary differential equation of the exceptional orthogonal polynomials of Laguerre type $X_1$ ( case I, II and III). The analogue of Rodrigues formulas for the general solution are constructed. The set of finite states from which the other states can be obtained algebraically is not unique but the vanishing arrow and diagonal arrow from the diagram of the 2-chain representations can be used to obtain minimal sets. These Rodrigues formulas are then exploited, not only to construct all the states (polynomial and non-polynomial), in a purely algebraic way, but also to obtain coefficients from the action of the ladder operators also in an algebraic manner. Those results are established via higher commutation relations related to the cubic Heisenberg Weyl algebra. The zero modes are associated with eigenstates but also generalised eigenstates.
\end{abstract}

\\
Email:  i.marquette@latrobe.edu.au
\bigskip
\noindent
%
%

\newpage

\tableofcontents

\newpage

\section{Introduction}

Exceptional orthogonal polynomials were discovered 15 years ago and since they appeared in different context of mathematical physics \cite{go09, go10, oda09, cq09}. It was pointed out \cite{cq09,oda09} that they were intimately related to the wavefunctions of several quantum isospectral and almost isospectral Hamiltonian deformation obtained via supersymmetric quantum mechanics (SUSYQM) \cite{dar88,crum55, kre57,adl94, jun96, bag01, fer99}. SUSYQM is also known as state-deleting, state adding, Krein-Adler and Darboux-Crum transformations. Later those Krein-Adler and Darboux-Crum transformations were used to provide a systematic approach for the classification of those exceptional orthogonal polynomials, then described by their codimensions \cite{fel09, cq11,oda13, ull13, ull14, gom16, gom20}. The classification was extended beyond the exceptional orthogonal polynomials (EOPs) associated with the classical orthogonal polynomials (Hermite, Laguerre and Jacobi) to Hahn and other types among the Askey-Wilson scheme of orthogonal polynomials. The classification, but also description of their properties is still an ongoing problem. In parallel, to those works which connect with one-dimensional quantum Hamiltonians it was pointed out \cite{mar13a} that Hermite EOPs $X_2$ was observed in context of superintegrable systems \cite{mar09} and that wider families of quantum superintegrable systems (i.e. multi-dimensional systems with more conserved quantities than degrees of freedom) can be constructed from knowledge of reccurence relations of EOPs. It was also illustrated that distinct set of ladder operators and related recurrence relations exist and can be exploited in various aspects such as algebraic descriptions via polynomial Heisenberg-Weyl algebras \cite{mar13b,mar13c,mar14}. 

However, the study of their associated differential equation is much more recent and only the example of the Hermite EOP $X_2$ was studied via different approaches. The polynomials part of the general solutions was studied via direct construction in \cite{ses10} and analysis of the series solution from the Frobenius approach \cite{car08}. To our knowledge this is the only known example of differential equations associated with EOPs for which series solution was investigated and apply to construct EOPs. Some further construction within the SUSYQM framework pointed out that wider set of states (non-physical/non-polynomials) in particular associated with the gap in the sequence of orthogonal polynomials could play a role describing the different properties. Some generalised eigenstates were also obtained \cite{car16} and it was pointed out they are useful in context of the underlying quantum system. However, the first study of the general solutions via series solutions was only performed very recently \cite{cha20}. In recent work, further insight into different formula for the general solutions as combination of generalised hypergeometric, connection to confluent Heun equations, recurrence and Rodrigues formula \cite{mar23}. It was also pointed out how linearly independent parts of the general solutions still form representations of the underlying polynomial Heisenberg-Weyl algebra and how the coefficients associated with the action of the generators can be obtained from different explicit formula in terms of generalised hypergeometric functions. It was discovered they form 2-chains type of representations.

This paper is devoted to present an entirely algebraic approach to this problem i.e. description of general solutions for the ordinary differential equations of given EOPs. The paper will consider an important case of Laguerre exceptional orthogonal polynomial that come in three different families I, II and III. We will rely on the factorization property of the ladder operators, construction of zero modes, commutator identities and induced type representations. This paper will also provide generalization of induced construction approach in context of cubic algebras and provide commutator identities that may be of interest for the study of such algebraic structures. 

The paper is organised as follow. In Section 2, we recall aspect of the construction of Darboux transformation for the three cases of seed functions. In Section 3, we present a construction of the ladder operators and cubic Heisenberg algebra. In Section 4, we present the induced representation construction and Rodrigues type formula. In Section 5, we present explicit construction for the type I and interpret as 2-chains representations for the cubic Heisenberg-Weyl algebra. In Section 6 and 7 we present explicit construction of 2 chains representations for the cubic Heisenberg-Weyl algebras of the EOPs Laguerre of type II and III repsectively.

\section{Darboux transformations and deformations of singular oscillator}

In this section, we recall several aspects of the singular oscillator and the deformed oscillator related to the Laguerre exceptional orthogonal polynomials $X_1$ \cite{cq09, cq11, mar13a,mar13b, mar14} in order to explicitly obtain the action related to 2-chain representations for two different types of ladder operators. We first consider the Hamiltonian of the singular oscillator

\begin{equation}
 H= - \frac{d^2}{dx^2} + x^2  +  \frac{l(l+1)}{x^2}.
\end{equation}
The eigenstates and energy spectrum are given by eq(\ref{states}) and eq(\ref{spectrum})
\begin{equation}
 \psi(\nu, \alpha,x)= x^{\alpha + \frac{1}{2}} e^{- \frac{1}{4} x^2} L_{\nu}^{\alpha}( \frac{1}{2} x^2 ),  \label{states}
\end{equation}
\begin{equation}
 E_{\nu}= 2 \nu + \alpha +1. \label{spectrum}
\end{equation} 

Using solution in terms of confluent hypergeometric functions with $l=\alpha - \frac{1}{2}$ it is possible to show that only three cases of seeds solution for Darboux-Crum transformaton ( state adding supersymmetric quantum mechanics in this case) allow for regular Hamiltonians i.e. Hamiltonians that allow to connect the Hilbert space of square integrable wavefunctions of both partner Hamiltonians. If additional singularities are created by the transformation then the map between both superpartner is usually only formal. We then set $z=\frac{1}{2} x^2$ and obtain explicitly in (\ref{phiI}), (\ref{phiII}) and (\ref{phiIII})

\begin{equation}
 \phi_I= z^{\frac{1}{4}(2 \alpha +1)} e^{\frac{1}{2}z} L_{1}^{\alpha}(-z),\quad E_I= -(\alpha +2 +1) , \label{phiI}
\end{equation}
\begin{equation} 
 \phi_{II}= z^{-\frac{1}{4}(2 \alpha -1)} e^{-\frac{1}{2}z} L_{1}^{-\alpha}(z),\quad E_{II}=-(\alpha -2 -1) , \label{phiII}
\end{equation}
\begin{equation} 
 \phi_{III}= z^{-\frac{1}{4}(2 \alpha -1)} e^{\frac{1}{2}z} L_{1}^{-\alpha}(-z),\quad E_{III}=\alpha -2 -1 . \label{phiIII}
\end{equation} 

The construction of partner Hamiltonian is well known and the polynomial part of the solution of the corresponding ODE using Darboux operators. We take

\begin{equation}
 (- \frac{d^2}{dx^2} + V^{+}(x) ) \phi(x) = E \phi(x) , \label{scro}
\end{equation} 

if the energy $E$ is smaller than the ground state energy $E_0^{+}$ of $H^{(+)}$ then the seed solution is nodeless but unnormalizable. The superpotential is determined $q =- \frac{\phi'}{\phi}$. Then the potential can be expressed as eq(\ref{potv})
\begin{equation}
 V^{\pm}= q^2 \mp q' + E , \label{potv}
\end{equation} 
and the corresponding Hamiltonian $H^{(+)}$ and $H^{(-)}$ as (\ref{hphm})
\begin{equation}
 H^{(+)} = A^{\dagger} A +E ,\quad H^{(-)} = A A^{\dagger} + E, \label{hphm}
\end{equation} 
in terms of the supercharges $A$ and $A^{\dagger}$. The supercharges (or intertwining operators) take the form
\begin{equation}
 A= \frac{d}{dx} + q,\quad  A^{\dagger}=- \frac{d}{dx} + q. \label{darb}
\end{equation} 

Here we have three type of seed functions $\phi_{I}$, $\phi_{II}$ and $\phi_{III}$. They lead to different superpotential $q_I$, $q_{II}$ and $q_{III}$ and distinct Hamiltonians $H_{I}^{(-)}$,  $H_{II}^{(-)}$ $H_{III}^{(-)}$. Here we provide details at the case $m=1$ and corresponding $H_{I}^{(-)}$, $H_{II}^{(-)}$ $H_{III}^{(-)}$. Here the initial Hamiltonian is the same in the three cases $H^{+}=H_{I}^{(+)}= H_{II}^{(+)}= H_{III}^{(+)}$. The Hamiltonian and eigenstates are given by
\begin{equation}
  H^{(+)} = - \frac{d^2}{dx^2} + \frac{x^2}{4} + \frac{\alpha^2 - \frac{1}{4}}{x^2} ,\quad \psi_{p}^{(+)}= e^{- \frac{1}{4} x^2} x^{\frac{1}{2} + \alpha} L_{\nu}^{\alpha}(\frac{x^2}{2}). 
\end{equation} 

This provides the following Hamiltonians $H_I^{(-)}$, $H_{II}^{(-)}$ and $H_{III}^{(-)}$  ( denoted as $H_J^{(-)}$ in (\ref{hm}) )
\begin{equation}
 H_{J}^{(-)}= A_{J} A_{J}^{\dagger} + E_J =  -\frac{d^2}{dx^2} + \frac{1}{4} x^2  + \frac{3 +4 \alpha (2 +\alpha)}{4x^2} -1 + \frac{8 x^2}{(F_J(x))^2} - \frac{4}{F_J(x)}, \label{hm}
\end{equation} 
with $F_J(x)$ where $J=\{I,II,III\}$ where
\begin{equation}
F_I(x)=2+ x^2 +2\alpha,\quad F_{II}(x)=-2+ x^2 +2\alpha,\quad F_{III}(x)=-2- x^2 +2\alpha.
\end{equation}
We can demonstrate using the polynomial solution of $H_I^{(+)}$ that
\begin{equation}
 \psi_{p,J}^{(-)}= A_{J} \psi_{p}^{(+)},\quad H_{J}^{(-)} \psi_{p,J}^{(-)} = E_{J} \psi_{p,J}^{(-)} ,\quad J=\{I,II,III\},
 \end{equation} 
with energy $E_{\nu,J}= 2\nu +1+\alpha$ for each cases and with the missing state
\begin{equation}
  \psi_{o,J}^{(-)}= \frac{1}{\phi_J},\quad J=\{I,II,III\},
\end{equation}  
with energy $E_0=-3-\alpha$. We will use the Schrodinger form given by eq(\ref{hm}), transformation to standard form can be made but as those transformation (often called similarity/gauge transformations) preserve spectrum, ladder and generating spectrum algebra they are equivalent setting. 

\section{Ladder operators and polynomial Heisenberg-Weyl algebras}

The initial Hamiltonian is the singular oscillator in the three cases $H_{I}^{(+)}=H_{II}^{(+)}=H_{III}^{(+)}$. Here the Darboux transformations have been established such that the initial Hamiltonian coincide in the three cases. The ladder operators for the singular oscillator (the initial Hamiltonian) on which the Darboux transformation is applied are. Also by construction the ladder operators are the same for all initial Hamiltonians $a=a_{I}=a_{II}=a_{III}$ and  $a^{+}=a_{I}^{+}=a_{II}^{+}=a_{III}^{+}$ and given by (\ref{a}) and (\ref{ad})
\begin{equation}
 a= \frac{1}{4}(2 \partial_x^2 +2 x \partial_x + \frac{1}{2} x^2 -2 \frac{l(l+1)}{x^2} +1), \label{a}
\end{equation}
\begin{equation}
 a^{+}= \frac{1}{4}(2 \partial_x^2 - 2 x \partial_x + \frac{1}{2} x^2 -2 \frac{l(l+1)}{x^2} -1). \label{ad}
\end{equation} 
 They lead to the Heisenberg-Weyl algebra given by (\ref{hwa})
\begin{equation}
 [H^{(+)},a]=-2 a, [H^{(+)},a^{\dagger}]=2 a^{\dagger} , [a,a^{\dagger}]= H^{(+)} . \label{hwa}
\end{equation} 
We will now rely on the three type of Darboux transformation and ladder of the singular oscillator to construct a set of ladder operator for the deformed Hamiltonians of type I,II and III. The standard ladder operators of fourth order of the $H_{I}^{(-)}$, $H_{II}^{(-)}$ and $H_{III}^{(-)}$ are respectively (\ref{Bj})
\begin{equation}
 B_{J}= A_{J} a A_{J}^{\dagger} ,\quad B_{J}^{\dagger}= A_{J} a A_{J}^{\dagger}, \quad J=\{I,II,III\}. \label{Bj}
\end{equation}
They satisfy cubic Heisenberg-Weyl algebras given respectively by (\ref{chwa1}) and (\ref{chwa2}) with (\ref{chwaI}), (\ref{chwaII}) and (\ref{chwaIII})
\begin{equation}
 [H_{J}^{(-)},B_{J}]=-2 B_{J} , [H_{J}^{(-)},B_{J}^{\dagger}]=2 B_{J}^{\dagger}, \label{chwa1}
\end{equation} 
\begin{equation}
  [B_{J},B_{J}^{\dagger}]= S^{J}(H_{J}^{(-)}),\quad J=\{I,II,III\} ,\label{chwa2}
 \end{equation} 
\begin{equation}
 S^{I}(H_{I}^{(-)})= (1+2 H_{I}^{(-)} -\alpha)(1+H_{I}^{(-)}+\alpha)(3+H_{I}^{(-)}+\alpha), \label{chwaI}
 \end{equation} 
\begin{equation} 
 S^{II}(H_{II}^{(-)})= (-1+2H_{II}^{(-)}-\alpha)(-3+H_{II}^{(-)}+\alpha)(-1+H_{II}^{(-)}+\alpha), \label{chwaII}
\end{equation} 
\begin{equation} 
S^{III}(H_{III}^{(-)})= (1+H_{III}^{(-)}-\alpha)(3+H_{III}^{(-)}-\alpha)(1+2 H_{III}^{(-)}+\alpha). \label{chwaIII}
\end{equation} 
As the cubic Heisenberg-Weyl algebra is obtained via explicit differential operator, there are other type of relations, among them the closure or product relations given by (\ref{prodI}), (\ref{prodII}) and (\ref{prodIII})
\begin{equation}
 B_{I}^{\dagger} B_{I}=R^{I}(H_{I}^{(-)})= \frac{1}{4} (H_{I}^{(-)}-1-\alpha)(H_{I}^{(-)}-1+\alpha)(H_{I}^{(-)}+1+\alpha)(H_{I}^{(-)}+4+\alpha), \label{prodI}
\end{equation}
\begin{equation}
 B_{II}^{\dagger} B_{II} = R^{II}(H_{II}^{(-)})=  \frac{1}{4} ( H_{II}^{(-)}-1 -\alpha)( H_{II}^{(-)}-5+\alpha)(H_{II}^{(-)}-3 +\alpha)(H_{II}^{(-)}-1+\alpha), \label{prodII}
\end{equation} 
\begin{equation}
 B_{III}^{\dagger} B_{III} = R^{III}(H_{III}^{(-)})= \frac{1}{4} ( H_{III}^{(-)} -\alpha -1)( H_{III}^{(-)}+1-\alpha)(H_{III}^{(-)}+3 -\alpha)(H_{III}^{(-)}-1+\alpha). \label{prodIII}
 \end{equation}
 Those closure relations also provide information on the eigenvalues of the zero modes of the lowering operators. Other polynomials associated with $B_{I} B_{I}^{\dagger}$,  $B_{II} B_{II}^{\dagger}$ and $B_{III} B_{III}^{\dagger}$ and given by $R^{I}(H_{I}^{(-)}+2)$, $R^{II}(H_{II}^{(-)}+2)$ and $R^{III}(H_{III}^{(-)}+2)$ are related with eigenvalues of the zero modes of the raising operator.

\section{Algebraic constructions and representations}
In this section, we develop commutator identities in order to define the induced representations, those are apriori infinite dimensional. However, they will be used to construct the representation associated with the two linearly independent solution of the underlying ordinary differential equations. Those will take the form of 2 chains of states interconnected via ladder operators, the formula of this section will be applied unless they lead to another zero modes which then will be use as new starting point. From mathematical point of view, the coefficient of the action of the generators will form a generalisation of Rodrigues type formula which are known for classical orthogonal polynomials. We consider the following generic cubic Heisenberg-Weyl algebra (\ref{calg1}), (\ref{calg2}) and (\ref{calg3}) with $\{H,c,c^{\dagger},\mathbbm{1}\}$
\begin{equation}
 [H,c]= -a c , \label{calg1}
\end{equation} 
\begin{equation}
 [H,c^{+}]=a c^{+}, \label{calg2}
\end{equation}
\begin{equation} 
 [c,c^{+}]= b_3 H^3 + b_2 H^2 + b_1 H  + b_0. \label{calg3}
\end{equation} 
Using the relations (\ref{calg1}), (\ref{calg2}) and (\ref{calg3}) we can demonstrate the commutator identity (\ref{com1})
\[ [c,(c^{+})^k]= (c^{+})^{k-1} f_n(H) = (c^{+})^{k-1} (  (a_0 + a_1 n) H^3 + ( a_2 + a_3 n + a_4 n^2 ) H^2 + ( a_5 + a_6 n + a_7 n^2 + a_8 n^3 ) H \]
\begin{equation}
 + (a_9 + a_{10} n + a_{11} n^2 + a_{12} n^3 + a_{13} n^4 ). \label{com1}
\end{equation} 
The coefficients $a_i$ are given in terms of the structure constants
\begin{equation}
 a_0 =0,\quad a_1 = b_3 ,\quad a_3 = \frac{1}{2} (-3 a b_3 +2 b_2 ) ,
\end{equation} 
\[ a_4 = \frac{3}{2} a b_3 ,\quad a_5 =0,\quad a_6 = \frac{1}{2} ( a^2 b_3 +2 b_1 - 2 a b_2 ), \]
\[ a_7 = \frac{1}{2} (-3 a^2 b_3 +2 a b_2 ) ,\quad a_8 = a^2 b_3 ,\quad a_9 =0, \]
\[ a_{10}= \frac{1}{6} ( 6 b_0 -3 a b_1 +a^2 b_2 ) ,\quad a_{11} = \frac{1}{4}(a^3 b_3 +2 a b_1 -2 a^2 b_2), \]
\[ a_{12} = \frac{1}{6} (-3 a^3 b_3 + 2 a^2 b_2 ),\quad a_{13} = \frac{a^3}{4} b_3. \]
In similar way, we can also demonstrate the following commutator identity (\ref{com2})
\[ [c^{+},(c)^k]= (c)^{k-1} g_n(H) = (c)^{k-1} (  (d_0 + d_1 n) H^3 + ( d_2 + d_3 n + d_4 n^2 ) H^2 + ( d_5 + d_6 n + d_7 n^2 + d_8 n^3 ) H \]
\begin{equation}
 + (d_9 + d_{10} n + d_{11} n^2 + d_{12} n^3 + d_{13} n^4 ). \label{com2}
\end{equation} 
The coefficients $d_i$ are given in terms of the structure constantw
\begin{equation}
 d_0 =0,\quad d_1=b_3 ,\quad d_2=0 ,\quad d_3= -\frac{3}{2} a b_3,\quad d_4= \frac{3a}{2} b_3 ,
\end{equation} 
\[ d_5 =0,\quad d_6= \frac{1}{2} (-a^2 b_3 -2 b_1 -2 a b_2) ,\quad d_7=\frac{1}{2} (3 a^2 b_3 +2 a b_2), \]
\[ d_8 =-a^2 b_3 ,\quad d_9 =0,\quad d_{10}= \frac{1}{6} (-6 b_0 -3 a b_1 -a^2 b_2 ),\]
\[ d_{11}= \frac{1}{4} ( a^3 b_3 + 2 a b_1 + 2 a^2 b_2 ),\quad d_{12} = \frac{1}{6} (-3 a^3 b_3 -2 a^2 b_2 ) ,\quad d_{13} = \frac{1}{4} a^3 b_3. \]
The identities established in formula (\ref{com1}) and (\ref{com2}) will be useful in context of constructing induced representations.

\subsection{Induced representations: the case of one infinite chain of states}
Here we consider the simplest types of induced representations (lowest and highest) corresponding to only one chain. However, they will be employed as building blocks on more complicated pattern of representations (2-chains cases) that we will investigate in later sections. We first consider a state with the following action of the lowering generator $c$ and $H$
\begin{equation}
c \chi_0=0, \quad H \chi_0 = \lambda \chi_0. \label{low1}
\end{equation}
The induced states are obtained by the action of the raising operator $c^{\dagger}$
\begin{equation}
\chi_n = (c^{\dagger})^n \chi_0 . \label{low2}
\end{equation}
Then we obtain the following action of the generators $c$ and $c^{\dagger}$ on the states $\chi_n$ as direct consequences 
\begin{equation}
c^{\dagger} \chi_n = \chi_{n+1},\quad c \chi_n= f_n(\lambda) \chi_{n-1} .\label{low3}
\end{equation}
where $f_n(\lambda)$ take different form for the three types I, II and III. The eq(\ref{low3}) constitute the action of generators on the infinite dimensional representations, of lowest weight type. Alternatively, we consider state such as
\begin{equation}
c^{\dagger} \xi_0=0,\quad H \xi_0 = \rho \xi_0, \label{hig1}
\end{equation}
with the induced states of the following form
\begin{equation}
\xi_n = (c)^n \xi_0. \label{hig2}
\end{equation}
This provides the following result for the action of $c$ and $c^{\dagger}$ on $\xi_n$ 
\begin{equation}
c \xi_n = \xi_{n+1},\quad c^{\dagger} \xi_n = g_n(\rho) \xi_{n-1}, \label{hig3}
\end{equation}
where $g_n(\rho)$ will take different form for the cases I,II and III.
This provides an infinite dimensional representation, analog of a highest weight. The two types of representations play an important roles. However, as it will be discussed in coming sections raising and lowering operators of higher order allow multiple zero modes which can be taken as starting point of a chain. They are all then connected in complicated way. The formula given by (\ref{low3}) and (\ref{hig3}) will still be useful to generate the action for broader types of representations. 

\subsection{ Function $f_n$ and $g_n$ for the cases of EOPs I, II and III}

Here we present the functions $f_n$ and $g_n$ for the different cases that are in formula (\ref{com1}), (\ref{com2}), (\ref{low3}) and (\ref{hig3}). We denote those functions as $f_n^{I}(H_{I}^{(-)})$, $f_n^{II}(H_{II}^{(-)})$, $f_n^{III}(H_{III}^{(-)})$, $g_n^{I}(H_{I}^{(-)})$, $g_n^{II}(H_{II}^{(-)})$ and $g_n^{III}(H_{III}^{(-)})$. They are given by the formula below (\ref{fn1}), (\ref{fn2}), (\ref{fn3}), (\ref{gn1}), (\ref{gn2}) and (\ref{gn3})

\begin{equation}
f_n^{I}(H_{I}^{(-)})= 2n (H_{I}^{(-)})^3 +3 n(2 + n + \alpha) (H_{I}^{(-)})^2 + 2n ( 1+ n(3+n)) H_{I}^{(-)} \label{fn1}
\end{equation}
\[ - \frac{n}{2} + \frac{n^4}{2} + n^3 (2 + \alpha) + \frac{1}{2} n^2 (2 +3 \alpha) - \frac{1}{2} \alpha (3 +2 \alpha (3 + \alpha)), \]
\[g_n^{I}(H_{I}^{(-)}) = -2 n (H_{I}^{(-)})^3 + 3 n (-4 +n -\alpha) (H_{I}^{(-)})^2 + ( - 2 n (10+(n-6)n) +3 (n-3) n \alpha) H_{I}^{(-)}\]
\begin{equation}
\- \frac{19n}{2} + \frac{n^4}{2} - n^3 (4 + \alpha) + \frac{1}{2} n^2 (20 + 9 \alpha) + \frac{1}{2} n \alpha (-9 +2 \alpha (3 +\alpha)) , \label{gn1}
\end{equation}
\begin{equation}
f^{II}_n( H_{II}^{(-)}) = 2 n (H_{II}^{(-)})^3 + 3n (-4 + n + \alpha) (H_{II}^{(-)})^2 + (  2n ( 10 + (n-6)n ) +3 (n-3) n \alpha ) H_{II}^{(-)}  \label{fn2}
\end{equation}
\[ - \frac{19n}{2} + \frac{n^4}{2} + \frac{1}{2} n^2 (20 - 9 \alpha) + n^3 (\alpha-4) + \frac{1}{2} n \alpha ( 9 -2 ( \alpha -3) \alpha), \]
\begin{equation}
g_n^{II}(H_{II}^{(-)}) -2 n (H_{II}^{(-)})^3 + 3 n (2 + n -\alpha) (H_{II}^{(-)})^2 + ( -2 n ( 1+ n ( 3+n)) + 3 n (1+n)\alpha) H_{II}^{(-)}  \label{gn2}
\end{equation}
\[ - \frac{n}{2} + \frac{n^4}{2} + \frac{1}{2} n^2 (2-3\alpha) -n^3 (\alpha-2) + \frac{1}{2} n \alpha ( 3 +2 (-3 +\alpha) \alpha), \]
\begin{equation}
f_n^{III}(H_{III}^{(-)})= 2 n (H_{III}^{(-)})^3 + 3n (2+n -\alpha) (H_{III}^{(-)})^2 + 2n ( 1+ n (3+n))(H_{III}^{(-)})  \label{fn3}
\end{equation}
\[ - \frac{n}{2} + \frac{n^4}{2} + \frac{1}{2} n^2 (2-3\alpha) -n^3(-2+\alpha) + \frac{1}{2} n \alpha (3 +2 (-3 + \alpha)\alpha), \]
\begin{equation}
g_n^{III}(H_{III}^{(-)})= -2 n (H_{III}^{(-)})^3 + 3 n ( -4 +n + \alpha) (H_{III}^{(-)})^2  \label{gn3}
\end{equation}
\[ - \frac{19n}{2} + \frac{n^4}{2} + \frac{1}{2} n^2 (20 -9 \alpha) + n^3 (\alpha -4) + \frac{1}{2} n \alpha ( 9 - 2 (\alpha -3)\alpha). \]

In the cases we will consider they will be evaluated on a given initial states ( from which others are induced ) and then $f_n^{I}(H_{I}^{(-)})$, $f_n^{II}(H_{II}^{(-)})$, $f_n^{III}(H_{III}^{(-)})$, $g_n^{I}(H_{I}^{(-)})$, $g_n^{II}(H_{II}^{(-)})$ and $g_n^{III}(H_{III}^{(-)})$ will be functions of $n$ the label (or quantum number). The energy spectrum will be linear in the quantum number in all cases. Those expressions (\ref{fn1} to (\ref{gn3}) will be evaluated at some explicit states that are used to generate other states and then those expression will only depends on parmameter $\alpha$ and $n$.

\section{Representations for the ODE of the EOPs Laguerre type I }

In order to achieve the construction of the 2-chains representations, We need to obtain first the pattern of zero modes for the lowering and raising operators, respectively $B_{I}$ and $B_{I}^{\dagger}$. Those sets of states will be later used to build all the other states via induced representations. However, due to the presence of generalised states among the zero modes we will need to add certain states into the set that are eigenstates.

\subsection{Zero modes of the ladder operators Case I}
The lowering operator $B_I= A_I a A_I^{\dagger}$, a fourth order operator, can be factorized into 3 distinct operators, there are three cases to investigate which provide 1, 2 and 1 solutions respectively. Instead of considering directly the vanishing action of $B_I$ for some state $\psi$ as
\begin{equation}
B_I \psi =0,
\end{equation}
we decompose the problem into three subcases ( denoted cases 1, 2 and 3 ). The cases 1, 2 and 3 consist in respectively (\ref{case1I}), (\ref{case2I}) and (\ref{case3I})
\begin{equation}
A_I^{\dagger} \psi =0, \label{case1I}
\end{equation}
\begin{equation}
a A_I^{\dagger} \psi =0, \label{case2I}
\end{equation}
\begin{equation}
A_I a A_I^{\dagger} \psi =0. \label{case3I}
\end{equation}
For the case 2, this means in fact we can proceed in using two coupled equations of the form given by eq(\ref{case2Ib})
\begin{equation}
a_I \tilde{\psi}=0,\quad A_I^{\dagger} \psi= \tilde{\psi}. \label{case2Ib}
\end{equation}
For the case 3, this means that in fact we can proceed into three coupled equations given as eq(\ref{case3Ib})
\begin{equation}
A_I \doubletilde{$\psi$}=0,\quad a \tilde{\psi} = \doubletilde{$\psi$},\quad A_I^{\dagger} \psi= \tilde{\psi} .\label{case3Ib}
\end{equation}
Then we obtain from case one given by (\ref{case1I}) the state $\psi_1$, from case two given by (\ref{case2Ib}) the states $\psi_2$ and $\psi_3$, and from case 3 related to (\ref{case3Ib}) we get the state $\psi_4$. This is achieved  by solving directly the related ODE. They take the following form
\begin{equation}
\psi_1 = \frac{e^{-\frac{x^2}{4}} x^{-\frac{1}{2} -\alpha}}{2+x^2 +2 \alpha},\quad \psi_2 = \frac{e^{-\frac{x^2}{4}} x^{-\frac{1}{2} -\alpha} (-x^4 -4 x^2 (1+\alpha))}{4(2+x^2 +2 \alpha)},
\end{equation}
\[ \psi_3 = e^{-\frac{x^2}{4}} x^{-\frac{1}{2}-\alpha} \frac{(-x^{4+2\alpha} -2 x^{2+2\alpha}(2+\alpha))}{4\alpha (2+\alpha)(2+x^2 +2\alpha)},\quad \psi_4 =\frac{-2 (-1)^{-\alpha} e^{\frac{x^2}{4}} (-x)^{\alpha} x^{\frac{3}{2}}}{ 2+x^2 + 2 \alpha},\]
with the action of the Hamiltonian $H_I^{(-)}$ on $\psi_1$, $\psi_2$, $\psi_3$ and $\psi_4$ given as
\begin{equation}
H^{(-)}_I \psi_1= (-3-\alpha) \psi_1,\quad H^{(-)}_I \psi_2 = (1-\alpha) \psi_2 - 4 \alpha (1+\alpha) \psi_1, \label{hmpsi1234}
\end{equation}
\[ H^{(-)}_I \psi_3 = (1+\alpha) \psi_3,\quad H^{(-)}_I \psi_4 = (-1-\alpha) \psi_4. \]

\begin{figure}[h]
\[
\scalebox{0.9}{  \xymatrixcolsep{5pc}\xymatrix@1{
....  & \psi_{1}  \ar@/^{10mm}/[l]^{0}    &   ...   &     \psi_4  \ar@/^{10mm}/[l]^{0} &  ...  & \psi_3   \ar@/^{10mm}/[l]^{0}  & ...   }
  }
  \]
  \caption{Action of $B_I$ on zero modes.}  
\end{figure}
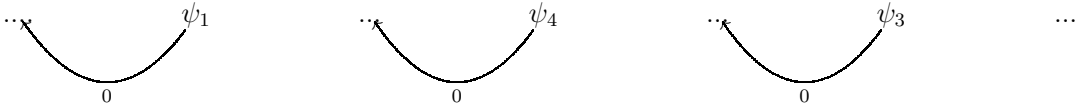

We now determine the pattern of zero modes for the vanishing action of  $B^{\dagger}_{I}$ on some state $\phi$. As the raising operator, a fourth order operator, is expressed as $B_I^{\dagger}= A_I a^{\dagger} A_I^{\dagger}$, there are three cases to investigate separately which provide 1, 2 and 1 solutions. Starting with the vanishing action of $B_I^{\dagger}$
\begin{equation}
B_I^{\dagger} \phi =0, 
\end{equation}
the three cases take the form of eq(\ref{case1It}), (\ref{case2It}) and (\ref{case3It}) respectively
\begin{equation}
A_I^{\dagger} \phi =0, \label{case1It}
\end{equation}
\begin{equation}
a^{\dagger} A_I^{\dagger} \phi =0, \label{case2It}
\end{equation}
\begin{equation}
A_I a^{\dagger} A_I^{\dagger} \phi =0. \label{case3It}
\end{equation}
The case 2, can be transformed into the following system of two coupled equations given by (\ref{case2Ibt})
\begin{equation}
a_I \tilde{\phi}=0,\quad A_I^{\dagger} \phi= \tilde{\phi}. \label{case2Ibt}
\end{equation}
and the case three reduce to the following set of three coupled equations given by (\ref{case3Ibt})
\begin{equation}
A_I \doubletilde{$\phi$}=0,\quad a \tilde{\phi} = \doubletilde{$\phi$},\quad A_I^{\dagger} \phi= \tilde{\phi}  .\label{case3Ibt}
\end{equation}
Then we obtain from case one $\phi_1$, from case two $\phi_2$ and $\phi_3$ and from case 3 $\phi_4$ and they take the explicit form (\ref{phi1234I})
\begin{equation}
\phi_1 = \frac{e^{-\frac{x^2}{4}} x^{-\frac{1}{2} -\alpha}}{2 +x^2 +2 \alpha},\quad \phi_2 = e^{\frac{x^2}{4}} x^{-\frac{1}{2} -\alpha} \frac{(x^2 + 2\alpha)}{2+x^2 +2 \alpha} , \label{phi1234I}
\end{equation}
\[\phi_3 = \frac{-e^{\frac{x^2}{4}} x^{\frac{3}{2} +\alpha}}{2\alpha(2+x^2+2\alpha)},\quad \phi_4 = \frac{(-1)^{-\alpha} e^{\frac{x^2}{4}} (-x)^{\alpha} x^{\frac{7}{2}} (4 +x^2 +4 \alpha)}{4(2+\alpha)(2+x^2 +2 \alpha)}. \]
The action of the Hamiltonian $H_I^{(-)}$ is then given by
\begin{equation}
H^{(-)}_I \phi_1= (-3-\alpha) \phi_1,\quad H^{(-)}_I \phi_2 = (\alpha-1) \phi_2 .
\end{equation}

We have $\phi_3=\frac{1}{4\alpha}\psi_4$ and $\phi_1=\psi_1$ which means we have two singlets. $\phi_4$ is a generalised eigenstates. Then overall we have $\psi_2$ and $\phi_4$ as generalised eigenstates and the eigenstates $\psi_1$, $\psi_4$, $\psi_3$ and $\phi_2$.

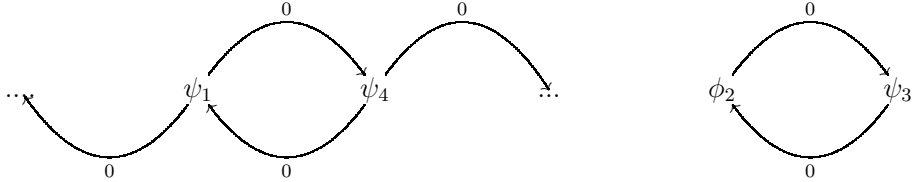
\begin{figure}[h]
\[
\scalebox{0.9}{  \xymatrixcolsep{5pc}\xymatrix@1{
....  & \psi_{1}  \ar@/^{10mm}/[r]^{0} \ar@/^{10mm}/[l]^{0}    &     \psi_4  \ar@/^{10mm}/[l]^{0}  \ar@/^{10mm}/[r]^{0}  &  ... & \phi_{2}  \ar@/^{10mm}/[r]^{0} &   \psi_3   \ar@/^{10mm}/[l]^{0}     }
  }
  \]
  \caption{Vanishing action of the ladder operators $B_I$ and $B_I^{\dagger}$ on zero modes.}  
\end{figure}

If $\alpha$ is not an integer then $B_I$ acting on $\phi_2$ will generate an infinite chains going below $\psi_1$. If $\alpha$ is an integer than the states will reach $\psi_1$ and coincide. We will now use a notation based, not on an index, but on the eigenvalue relative to the Hamiltonian H. The figure 2 and 3 provide further details on the first chain and structure of the zero modes.

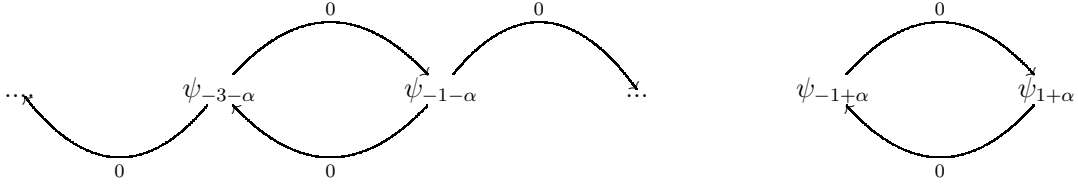
\begin{figure}[h]
\[
\scalebox{0.9}{  \xymatrixcolsep{5pc}\xymatrix@1{
....  & \psi_{-3-\alpha}  \ar@/^{10mm}/[r]^{0} \ar@/^{10mm}/[l]^{0}    &     \psi_{-1-\alpha}  \ar@/^{10mm}/[l]^{0}  \ar@/^{10mm}/[r]^{0}  &  ... &  \psi_{-1+\alpha}  \ar@/^{10mm}/[r]^{0} &   \psi_{1+\alpha}   \ar@/^{10mm}/[l]^{0}    }
}
  \]
  \caption{Action of the ladder operators and zero modes: Notation in terms of eigenvalues of the Hamiltonian}  
\end{figure}

\subsection{Induced representations: the first chain for type I}
This provide overall 4 eigenstates and two generalised eigenstates. The action of $B_I^{\dagger}$ lead to non trivial states and in fact an infinite chains of eigenstates can be generated (using (\ref{low1}) and (\ref{low2}) )
\begin{equation}
B_I \psi_{1+\alpha} =0,
\end{equation}
\begin{equation}
\psi_{1+\alpha +2n}= ( B_I^{\dagger} )^n \psi_{1+\alpha}. 
\end{equation}
Then from (\ref{low3}) and (\ref{fn1}) we get directly the following actions of $B_I$, $B_I^{\dagger}$ and $H$ on $\psi_{1+\alpha+2n}$
\begin{equation}
B_I^{\dagger} \psi_{1+\alpha +2n} = \psi_{1+\alpha + 2(n+1)},
\end{equation}
\begin{equation}
B_I \psi_{1+\alpha +2n} = f_n^{I}(1+\alpha) \psi_{1+\alpha+2(n-1)} ,
\end{equation}
\begin{equation}
H^{(-)} \psi_{1+\alpha+2n}= (1+\alpha+2n) \psi_{1+\alpha+2n}.
\end{equation}
In general the pattern of one-chains representations can be more complicated and even more for the 2-chains that will be developed using the two linearly independent solutions. In order to provide a complete algebraic description of the one-chain we need to identity appropriate state to apply the induced representation approach. The singlet and pattern and of zero modes greatly affect the choice of such states. We need to add in addition states using their Schrodinger equation

\begin{equation}
H^{(-)}_I \psi_{-5-\alpha}= (-5-\alpha) \psi_{-5-\alpha}.
\end{equation}

There are two linearly independent solution and among them the following which still involve a polynomial (up to exponential and rational factor).

\begin{equation}
\psi_{-5-\alpha}= \frac{e^{\frac{x^2}{4}} x^{\frac{3}{2}+\alpha} ( x^4 + 4 x^2 (1+\alpha) + 4 (1+\alpha)(2+\alpha))}{2+x^2 +2\alpha}.
\end{equation}
We let for the moment aside the other solution $\tilde{\psi}_{-5-\alpha}$. We have for $\psi_{-5-\alpha}$ that
\begin{equation}
B^{\dagger}_{I} \psi_{-5-\alpha} =0.
\end{equation}
This was a zero mode related to the generalized eigenstates $\phi_4$, this point out how in this context identifying all zero modes and relevant states to apply the induced construction may be less straightforward than Lie algebras. Then we define the states via
\begin{equation}
\psi_{-5-\alpha-2n}= (B_I)^{n} \psi_{-5-\alpha}.
\end{equation}
We also obtain the following action on the states
\begin{equation}
B_I \psi_{-5-\alpha-2n}= \psi_{-5-\alpha-2(n+1)},
\end{equation}
\begin{equation}
B_I^{\dagger} \psi_{-5-\alpha-2n}= g_n^{I}(-5-\alpha) \psi_{-5-\alpha -2(n-1)}.
\end{equation}
This complete the structure of the representation for the first chain.

\subsection{Induced representation the second chain}
We now turn to the construction of the second chain to form a wider indecomposable representations. We start with the already known states ( i.e. $\psi_1$, $\psi_3$, $\phi_2$, $\phi_3$ now denoted $\psi_{-3-\alpha}$, $\psi_{1+\alpha}$, $\psi_{-1+\alpha}$ and $\psi_{-1-\alpha}$ (including now the eigenstate $\psi_{-5-\alpha}$ )
\begin{equation}
\tilde{\psi}_{-3-\alpha} = \psi_{-3-\alpha} \int \frac{1}{\psi_{-3-\alpha}^2} dx ,
\end{equation}
\begin{equation}
\tilde{\psi}_{1+\alpha} = \psi_{1+\alpha} \int \frac{1}{\psi_{1+\alpha}^2} dx ,
\end{equation}
\begin{equation}
\tilde{\psi}_{-1+\alpha} = \psi_{-1+\alpha} \int \frac{1}{\psi_{-1+\alpha}^2} dx ,
\end{equation}
\begin{equation}
\tilde{\psi}_{-1-\alpha} = \psi_{-1-\alpha} \int \frac{1}{\psi_{-1-\alpha}^2} dx ,
\end{equation}
\begin{equation}
\tilde{\psi}_{-5-\alpha} = \psi_{-5-\alpha} \int \frac{1}{\psi_{-5-\alpha}^2} dx .
\end{equation}
Some of them can be connected with $\phi_4$ and $\psi_2$ which are not eigenstates but generalised eigenstates. We now present in Figure 4, the action of lowering (arrow going to the left) and raising operators ( arrow going to the right) for the 2-chains representations. This will come from explicit calculations of the action of the ladder operators on the states.


\begin{figure}[h]
\begin{equation*}
\xymatrix{
...  & \psi_{-5-\alpha} \ar@/^{5mm}/[r]^{0} \ar@/^{5mm}/[l] &  \psi_{-3-\alpha} \ar@/^{5mm}/[l]^{0} \ar@/^{5mm}/[r]^{0}    &  \psi_{-1-\alpha} \ar@/^{5mm}/[l]^{0} \ar@/^{5mm}/[r]^{0}  & ... & \psi_{-1+\alpha}  \ar@/^{5mm}/[r]^{0} \ar@/^{5mm}/[l]  & \psi_{1+\alpha} \ar@/^{5mm}/[l]^{0} \ar@/^{5mm}/[r]  &  ...    \\ \\ \\
...  & \tilde{\psi}_{-5-\alpha} \ar@/^{5mm}/[l]  \ar@/^{3mm}/[uuur]   &  \tilde{\psi}_{-3-\alpha}    \ar@/^{3mm}/[uuur] &  \tilde{\psi}_{-1-\alpha} \ar@/^{3mm}/[uuul] \ar@/^{5mm}/[r]  & ... & \tilde{\psi}_{-1+\alpha} \ar@/^{3mm}/[uuur] \ar@/^{3mm}/[uuul] & \tilde{\psi}_{1+\alpha} \ar@/^{5mm}/[r] \ar@/^{3mm}/[uuul] &  ...
}
\end{equation*}
  \caption{Action of the ladder on the two set of states forming the 2-chains representations}  
\end{figure}
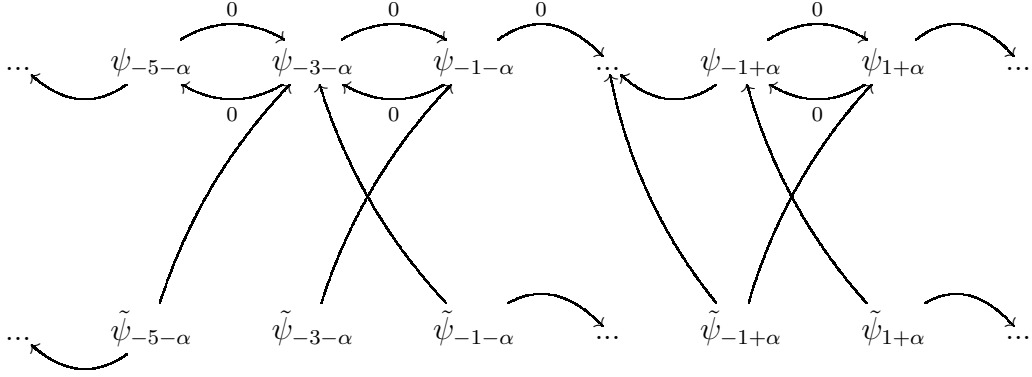

This allow to see that in order to induce other states of the two chains representations we need to take into account $\tilde{\psi}_{-5-\alpha}$, $\tilde{\psi}_{-1-\alpha}$ and $\tilde{\psi}_{1+\alpha}$. So far we have not connected the 2-chains formed by eigenstates of the Hamiltonians and the generalised eigenstates $\phi_4$ and $\psi_2$ ( setting $\phi_4=\hat{\psi}_{-5-\alpha}$
\begin{equation}
H_I^{(-)} \hat{\psi}_{-5-\alpha} = (-5-\alpha) \hat{\psi}_{-5-\alpha} - \frac{1}{2+\alpha} \psi_{-5-\alpha},
\end{equation}
\begin{equation}
B_I^{\dagger} \hat{\psi}_{-5-\alpha} =0,
\end{equation}
\begin{equation}
B_I \hat{\psi}_{-5-\alpha} \propto \psi_{-7-\alpha} + \psi_{-1-\alpha}.
\end{equation}
This means $\hat{\psi}_{-5-\alpha}$ is part of a generalised weight space with $\psi_{-5-\alpha}$. Setting as well ($\psi_2= \hat{\psi}_{-3-\alpha}$),  we have the following formula
\begin{equation}
H_I^{(-)} \hat{\psi}_{-3-\alpha} = (1-\alpha) \hat{\psi}_{-3-\alpha} -4 \alpha (1+\alpha) \psi_{-3-\alpha},
\end{equation}
\begin{equation}
B_I \hat{\psi}_{-3-\alpha} =0,
\end{equation}
\begin{equation}
B_I^{\dagger} \hat{\psi}_{-3-\alpha} \propto \psi_{3-\alpha} + \psi_{-1-\alpha}.
\end{equation}
Now we have clarified the relation between $\hat{\psi_{-5-\alpha}}$ and $\hat{\psi}_{-3-\alpha}$ with other states that are part of the two chains. We can generate other states via
\begin{equation}
\tilde{\psi}_{-5-\alpha-2n} = (B_I)^n \tilde{\psi}_{-5-\alpha},
\end{equation}
\begin{equation}
\tilde{\psi}_{-1-\alpha+2n} = (B_I^{\dagger})^n \tilde{\psi}_{-1-\alpha},
\end{equation}
\begin{equation}
\tilde{\psi}_{1+\alpha+2n} = (B_I^{\dagger})^n \tilde{\psi}_{1+\alpha}.
\end{equation}
As we have the following relation between $\tilde{\psi}_{1+\alpha}$ and $\psi_{-1+\alpha}$
\begin{equation}
B_I \tilde{\psi}_{1+\alpha} \propto \psi_{-1+\alpha} .
\end{equation}
We then calculate directly that
\begin{equation}
  B_I \tilde{\psi}_{1+\alpha+2n}= B_I (B_I^{\dagger})^n \tilde{\psi}_{1+\alpha}  
\end{equation}
\[ = f_n^{I}(1+\alpha) \tilde{\psi}_{1+\alpha+2(n-1)} + (B_I^{\dagger})^n \psi_{-1+\alpha} = f_n^{I}(1+\alpha) \tilde{\psi}_{1+\alpha+2(n-1)},   \]
as $B_I^{\dagger} \psi_{-1+\alpha}=0$. We also have the following relations
\begin{equation}
B_I \tilde{\psi}_{-1-\alpha}  \propto \psi_{-3-\alpha},\quad B^{\dagger} \psi_{-3-\alpha} =0,
\end{equation}
and obtain via direct calculation that the action is
\begin{equation}
 B_I \tilde{\psi}_{-1-\alpha+2n} =  f_n^{I}(-1-\alpha)  \tilde{\psi}_{-1-\alpha+2(n-1)} .
\end{equation}
Finally, we have the following action of $B_I^{\dagger}$ and $B_I$ on $\tilde{\psi}_{-5-\alpha}$ and $\psi_{-3-\alpha}$
\begin{equation}
B_I^{\dagger} \tilde{\psi}_{-5-\alpha} \propto \psi_{-3-\alpha},\quad B_I \psi_{-3-\alpha}=0.
\end{equation}
This allows as in other cases to obtain via formula of section 4 the action and explicit coefficient algebraically
\begin{equation}
 B_I^{\dagger} \tilde{\psi}_{-5-\alpha-2n}= B_I^{\dagger} B_I^n  \tilde{\psi}_{-5-\alpha}  = g_n^{I}(-5-\alpha )\tilde{\psi}_{-5-\alpha-2(n-1)}.
\end{equation}
This complete the description of the 2-chains representations and coefficients of the action of the generatorsa of the cubic Heisenberg-Weyl algebra. The resultds are summarized as well in Figure 4.

\section{Explicit construction of 2-chains representations for the case II}

\subsection{Zero modes of the ladder operators Case II}

Applying similar approch of breaking down solving the fourth order differential equations for the constraints of zero modes into three different cases as the operator factorise we also can obtain and classify the zero via their weights (eigenvalue of the Hamiltonian). The zero modes of the annihilation operators $B_{II}$ now take the form eq(\ref{psi1234II})
\begin{equation}
 \psi_1= \frac{e^{\frac{x^2}{4}} x^{-\frac{1}{2} +\alpha} }{-2+ x^2 +2\alpha} ,\quad  \psi_2= \frac{e^{\frac{-x^2}{4}} x^{\frac{3}{2} -\alpha} }{-2+ x^2 +2\alpha} , \label{psi1234II}
\end{equation}
\[ \psi_3= \frac{e^{-\frac{x^2}{4}} x^{-\frac{1}{2} +\alpha} (x^2 +2 \alpha) }{2\alpha(-2+ x^2 +2\alpha)}   ,\quad  \psi_4= \frac{e^{-\frac{x^2}{4}} x^{\frac{7}{2} -\alpha} (x^2 -4 +4 \alpha) }{4(\alpha-2)(-2+ x^2 +2\alpha)},  \]
with action of the Hamiltonian $H^{(-)}_II$
\begin{equation}
   H_{II}^{(-)}\psi_1 = (3-\alpha) \psi_1  ,\quad   H_{II}^{(-)}\psi_2 = (1-\alpha) \psi_2 ,
\end{equation}
\[   H_{II}^{(-)}\psi_3 = (1+\alpha) \psi_3  ,\quad   H_{II}^{(-)}\psi_4 = (5-\alpha) \psi_4 + (4\alpha -4) \psi_2.  \]
The zero mode of the raising operator are then given by (\ref{phi1234II})
\begin{equation}
\phi_1=  \frac{e^{\frac{x^2}{4}} x^{-\frac{1}{2} +\alpha} }{(-2+ x^2 +2\alpha)}   ,\quad \phi_2=  \frac{e^{\frac{x^2}{4}} x^{\frac{3}{2} -\alpha} (-4 + x^2 +2 \alpha) }{2(\alpha-2)(-2+ x^2 +2\alpha)},  \label{phi1234II}
\end{equation}
\[ \phi_3=  \frac{e^{\frac{x^2}{4}} x^{\frac{3}{2} +\alpha} (4 - x^2 -4 \alpha) }{8\alpha(-2+ x^2 +2\alpha)}   ,\quad 
\phi_4=  \frac{2e^{-\frac{x^2}{4}} x^{\frac{3}{2} -\alpha} }{(-2+ x^2 +2\alpha)},   \]
and the action of the Hamiltonian on those zero mode is given by
\begin{equation}
   H_{II}^{(-)}\phi_1 = (3-\alpha) \phi_1, \quad  H_{II}^{(-)}\phi_2 = (\alpha-1) \phi_2, 
\end{equation}
\[   H_{II}^{(-)}\phi_3 = (-\alpha-1) \phi_3 + (2\alpha -2) \psi_1  ,\quad
   H_{II}^{(-)}\phi_4 = (-\alpha+1) \phi_4.   \]

Due to the fact we have generalised eigenstates ($\psi_4$, $\phi_3$) as zero modes and their associated states are not with same weight this means that some states need to be added directly by solving the Schrodinger equations to obtain polynomial type of solution for eigenvalue $-\alpha-1$ and $5-\alpha$. We look then directly for states $\chi_1$ and $\chi_2$ which satisfy

 \begin{equation}
   H_{II}^{(-)}\chi_1 = (5-\alpha) \chi_1 ,
\end{equation}
\begin{equation}
   H_{II}^{(-)}\chi_2 = (-\alpha-1) \chi_2 .
\end{equation}
Seeking polynomial type of solution only, we get
\begin{equation}
   \chi_1=   \frac{-e^{-\frac{x^2}{4}} x^{\frac{3}{2} -\alpha}  (x^4 + 4 x^2 (\alpha-1) +4 (\alpha-2)(\alpha-1) )  }{(-2+ x^2 +2\alpha)},
\end{equation}
\begin{equation}
   \chi_2 =   \frac{e^{\frac{x^2}{4}} x^{-\frac{1}{2} +\alpha}  (x^4 + 4 x^2 (\alpha-1) +4 (\alpha-1)\alpha )  }{(-2+ x^2 +2\alpha)}.
\end{equation}
We then unify the description and take the following notation  below. We also remove from the set $\psi_4$ and $\phi_3$) and take into account $\phi_1=\psi_1$, $\phi_4=2 \psi_2$, this gives
\[  \psi_1  \rightarrow \psi_{3-\alpha},\quad  \psi_2  \rightarrow \psi_{1-\alpha}, \quad  \psi_3  \rightarrow \psi_{1+\alpha} , \]
\[  \psi_1  \rightarrow \psi_{3-\alpha},\quad  \psi_2  \rightarrow \psi_{1-\alpha}, \quad  \psi_3  \rightarrow \psi_{1+\alpha} , \]
and
\[  \chi_1 \rightarrow \psi_{5-\alpha},\quad \chi_2 \rightarrow \psi_{-\alpha-1}  . \]
We then construct element of the second chains using integral formula and obtain explicitly
\begin{equation}
\tilde{\psi}_{-\alpha-1} = \psi_{-\alpha-1} \int \frac{1}{\psi_{-\alpha-1}^2} dx ,
\end{equation}
\begin{equation}
\tilde{\psi}_{-\alpha+1} = \psi_{-\alpha+1} \int \frac{1}{\psi_{-\alpha+1}^2} dx ,
\end{equation}
\begin{equation}
\tilde{\psi}_{-\alpha+3} = \psi_{-\alpha+3} \int \frac{1}{\psi_{-\alpha+3}^2} dx ,
\end{equation}
\begin{equation}
\tilde{\psi}_{-\alpha+5} = \psi_{-\alpha+5} \int \frac{1}{\psi_{-\alpha+5}^2} dx ,
\end{equation}
\begin{equation}
\tilde{\psi}_{\alpha-1} = \psi_{\alpha-1} \int \frac{1}{\psi_{\alpha-1}^2} dx ,
\end{equation}
\begin{equation}
\tilde{\psi}_{\alpha+1} = \psi_{\alpha+1} \int \frac{1}{\psi_{\alpha+1}^2} dx .
\end{equation}
The action of the ladder operators $B_{II}$ and $B_{II}^{\dagger}$ on the states is given in Figure 5 ( again the action of the lowering operators toward the left and action of the raising operator toward the right ).
\begin{figure}[h]
\begin{equation*}
\xymatrix{
...  & \psi_{-\alpha-1} \ar@/^{5mm}/[r]^{0} \ar@/^{5mm}/[l] &  \psi_{-\alpha+1} \ar@/^{5mm}/[l]^{0} \ar@/^{5mm}/[r]^{0}    &  \psi_{-\alpha+3} \ar@/^{5mm}/[l]^{0} \ar@/^{5mm}/[r]^{0}   & \psi_{-\alpha+5}  \ar@/^{5mm}/[l]^{0} \ar@/^{5mm}/[r]  & ...  & \psi_{\alpha-1} \ar@/^{5mm}/[l] \ar@/^{5mm}/[r]^{0}  &   \psi_{\alpha+1}  \ar@/^{5mm}/[l]^{0} \ar@/^{5mm}/[r] &   ...    \\ \\ \\
...  & \tilde{\psi}_{-\alpha-1} \ar@/^{5mm}/[l]  \ar@/^{3mm}/[uuur]   &  \tilde{\psi}_{-\alpha+1}    \ar@/^{3mm}/[uuul] \ar@/^{5mm}/[r]^{0}  &  \tilde{\psi}_{-\alpha+3} \ar@/^{3mm}/[uuul]  \ar@/^{3mm}/[uuur]   & \tilde{\psi}_{-\alpha+5} \ar@/^{3mm}/[r]  \ar@/^{3mm}/[uuul] & ... & \tilde{\psi}_{\alpha-1} \ar@/^{5mm}/[l] \ar@/^{3mm}/[uuur]  &  \tilde{\psi}_{\alpha+1} \ar@/^{3mm}/[uuul]  \ar@/^{3mm}/[r]  &  ...
}
\end{equation*}
  \caption{Action of the ladder }  
\end{figure}
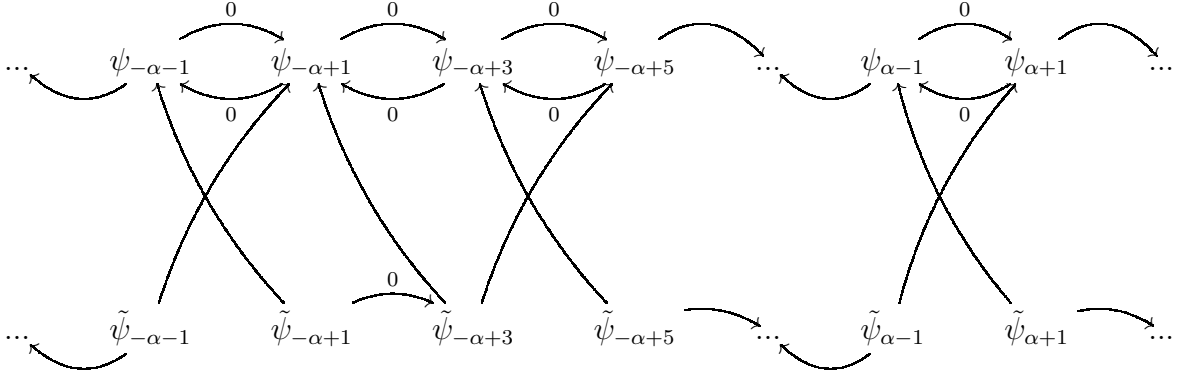

Explicitly the coefficient of the action that relate the two chains are given by the following formula 
\begin{equation}
B_{II}^{\dagger} \tilde{\psi}_{-\alpha+3}= \psi_{-\alpha+5},
\end{equation}
\begin{equation}
   B_{II} \tilde{\psi}_{-\alpha+3} = (4-4\alpha) \psi_{-\alpha+1},
\end{equation}
\begin{equation}
    B_{II} \tilde{\psi}_{-\alpha+5}=-\psi_{-\alpha+3},
\end{equation}
\begin{equation}
   B_{II}^{\dagger} \tilde{\psi}_{\alpha-1} =-4(\alpha-2)(\alpha-1) \alpha \psi_{\alpha+1} ,
\end{equation}
\begin{equation}
    B_{II} \tilde{\psi}_{\alpha+1} =4 (\alpha-2)(\alpha-1) \alpha \psi_{\alpha-1}.
\end{equation}
\begin{equation}
    B_{II}^{\dagger} \tilde{\psi}_{-\alpha-1} =- \psi_{-\alpha+1}.
\end{equation}
\begin{equation}
    B_{II} \tilde{\psi}_{-\alpha+1} = \psi_{-\alpha-1}.
\end{equation}
In order to get the complete action of all the states of the 2-chains that are infinite on both side, we need to filter the states as two types: the one that we need to add as they cannot be induced from other ones such as singlets and the one that allow to induce other ones. The one we need to add are

\[  \{\psi_{-\alpha+1}, \psi_{-\alpha+3}, \psi_{\alpha-1}, \tilde{\psi}_{-\alpha+1} , \tilde{\psi}_{-\alpha+3}, \tilde{\psi}_{\alpha-1} \} , \]
and the one to induce other ones are
\[  \{  \psi_{-\alpha-1}, \psi_{-\alpha+5}, \psi_{\alpha+1}, \tilde{\psi}_{-\alpha-1}, \tilde{\psi}_{-\alpha+5}, \tilde{\psi}_{\alpha+1} \} .   \]
The induced states are then given by
\begin{equation}
   \psi_{-\alpha-1-2n} = (B_{II})^n \psi_{-\alpha-1},
\end{equation}
\begin{equation}
   \psi_{-\alpha+5+2n} = (B_{II}^{\dagger})^n \psi_{-\alpha+5},
\end{equation}
\begin{equation}
   \psi_{\alpha+1+2n} = (B_{II}^{\dagger})^n \psi_{\alpha+1},
\end{equation}
\begin{equation}
   \tilde{\psi}_{-\alpha-1-2n} = (B_{II})^n \tilde{\psi}_{-\alpha-1},
\end{equation}
\begin{equation}
   \tilde{\psi}_{-\alpha+5+2n} = (B_{II}^{\dagger})^n \tilde{\psi}_{\alpha+5},
\end{equation}
\begin{equation}
   \tilde{\psi}_{\alpha+1+2n} = (B_{II}^{\dagger})^n \tilde{\psi}_{\alpha+1}.
\end{equation}
We then obtain the following action on the states 
\begin{equation}
    B_{II} \psi_{-\alpha-1-2n}=\psi_{-\alpha-1-2(n+1)},
\end{equation}
\begin{equation}
    B_{II}^{\dagger} \psi_{-\alpha-1-2n}= 
 g_n^{II}(-\alpha-1) \psi_{-\alpha-1-2(n-1)},
\end{equation}
\begin{equation}
    B_{II}^{\dagger} \psi_{-\alpha+5+2n}=\psi_{-\alpha+5+2(n+1)},
\end{equation}
\begin{equation}
    B_{II} \psi_{-\alpha+5+2n}= 
 f_n^{II}(-\alpha+5) \psi_{-\alpha+5+2(n-1)},
\end{equation}
\begin{equation}
    B_{II}^{\dagger} \psi_{\alpha+1+2n}=\psi_{\alpha+1+2(n+1)},
\end{equation}
\begin{equation}
    B_{II} \psi_{\alpha+1+2n}= 
 f_n^{II}(\alpha+1) \psi_{\alpha+1+2(n-1)}.
\end{equation}
Finally we also obtain 
\begin{equation}
    B_{II} \tilde{\psi}_{-\alpha-1-2n}=\tilde{\psi}_{-\alpha-1-2(n+1)},
\end{equation}
\begin{equation}
    B_{II}^{\dagger} \tilde{\psi}_{-\alpha-1-2n}= 
 g_n^{II}(-\alpha-1) \tilde{\psi}_{-\alpha-1-2(n-1)},
\end{equation}
\begin{equation}
    B_{II}^{\dagger} \tilde{\psi}_{-\alpha+5+2n}=\tilde{\psi}_{-\alpha+5+2(n+1)},
\end{equation}
\begin{equation}
    B_{II} \tilde{\psi}_{-\alpha+5+2n}= 
 f_n^{II}(-\alpha+5) \tilde{\psi}_{-\alpha+5+2(n-1)},
\end{equation}
\begin{equation}
    B_{II}^{\dagger} \tilde{\psi}_{\alpha+1+2n}=\tilde{\psi}_{\alpha+1+2(n+1)},
\end{equation}
\begin{equation}
    B_{II} \tilde{\psi}_{\alpha+1+2n}= 
 f_n^{II}(\alpha+1) \tilde{\psi}_{\alpha+1+2(n-1)}.
\end{equation}
This complete as well the description of the 2-chains representations for the case of Laguerre EOPs    of type II.

\section{Explicit construction of 2-chains representations for the case III}
Using again a similar approach to solve the fourth order differential equations for the existence of zero modes of the lowering and raising operators in terms of three cases based on the factorization of the operators. We then obtain 
\begin{equation}
\psi= \frac{ e^{-\frac{x^2}{x}} x^{-\frac{1}{2} +\alpha} }{2+x^2-2\alpha},\quad    \psi_2=  \frac{ e^{-\frac{x^2}{x}} x^{\frac{3}{2} -\alpha} (4+x^2 -2\alpha) }{2(-2+\alpha)(2+x^2-2\alpha)},
\end{equation}
\begin{equation}
   \psi_3=  \frac{ e^{-\frac{x^2}{x}} x^{\frac{3}{2} +\alpha} (4+x^2 -4\alpha) }{8 \alpha (2+x^2-2\alpha)},\quad     \psi_4 =  \frac{ -2 e^{-\frac{x^2}{x}} x^{\frac{3}{2} -\alpha} }{ (2+x^2-2\alpha)},
\end{equation}
with action of $H_{III}^{(-)}$ on those states given by
\begin{equation}
   H_{III}^{(-)}\psi_1 = (-3+\alpha)\psi_1 ,\quad   H_{III}^{(-)}\psi_2 = (1-\alpha)\psi_2 ,
\end{equation}
\begin{equation}
   H_{III}^{(-)} \psi_3 =  (1+\alpha) \psi_3 + (2+2 \alpha) \psi_1 ,\quad   H_{III}^{(-)}\psi_4 = (-1+\alpha)\psi_4.
\end{equation}
For the lowering operators we get
\begin{equation}
    \phi_1 =  \frac{  e^{-\frac{x^2}{x}} x^{-\frac{1}{2} +\alpha} }{ (2+x^2-2\alpha)},\quad   \phi_2 =  \frac{ - e^{\frac{x^2}{x}} x^{\frac{3}{2} -\alpha} }{ (2+x^2-2\alpha)},
\end{equation}
\begin{equation}
    \phi_3 =  \frac{ - e^{\frac{x^2}{x}} x^{-\frac{1}{2} +\alpha} (x^2 -2 \alpha) }{ 2\alpha (2+x^2-2\alpha)},\quad
    \phi_4 =  \frac{  e^{\frac{x^2}{x}} x^{\frac{7}{2} -\alpha} (4+x^2 -4 \alpha) }{ 4 (\alpha-2) (2+x^2-2\alpha)},
\end{equation}
with action of $H_{III}^{(-)}$
\begin{equation}
    H_{III}^{(-)}\phi_1 = (-3+\alpha)\phi_1,\quad     H_{III}^{(-)}\phi_2 = (\alpha-1)\phi_2
\end{equation}
\begin{equation}
    H_{III}^{(-)}\phi_3 = (-\alpha-1)\phi_3,\quad    H_{III}^{(-)}\phi_4 = (\alpha-5)\phi_4 + (-2+2\alpha) \psi_4.
\end{equation}
We have $\psi_4=2 \phi_2$ and $\psi_1=\phi_1$. We will need to add states with eigenvalues $\alpha+1$ and $\alpha-5$ as we have generalised eigenstates $\psi_3$ and $\phi_4$. We consider $\psi_1$, $\psi_2$, $\psi_4$, $\phi_3$. The related equations can be solved directly
\begin{equation}
H_{III}^{(-)} \chi_1= (\alpha +1) \chi_1,
\end{equation}
\begin{equation}
    \chi_1 = \frac{ e^{ - \frac{x^2}{4}  } x^{-\frac{1}{2} +\alpha} (x^4 - 4 x^2 (\alpha -1) +4 (\alpha -1)\alpha  }{2+x^2 -2 \alpha},
\end{equation}
\begin{equation}
H_{III}^{(-)} \chi_2= (\alpha -5) \chi_2,
\end{equation}
\begin{equation}
    \chi_2 = \frac{ e^{  \frac{x^2}{4}  } x^{\frac{3}{2} -\alpha} (x^4 - 4 x^2 (\alpha -1) + 4 (\alpha -2)(\alpha -1) }{2+x^2 -2 \alpha}.
\end{equation}
We denote the states by using their eigenvalues
\[  \psi_1 \rightarrow \psi_{\alpha-3} ,\quad \psi_2 \rightarrow  \psi_{-\alpha+1},\quad \psi_4 \rightarrow \psi_{\alpha-1} ,\quad  \phi_3 \rightarrow  \psi_{-\alpha-1} ,\quad    \]
\[ \chi_2  \rightarrow  \psi_{\alpha-5} ,\quad \chi_1 \rightarrow \psi_{\alpha_1}. \]
We construct the second chains of states as
\begin{equation}
\tilde{\psi}_{-\alpha-1} = \psi_{-\alpha-1} \int \frac{1}{\psi_{-\alpha-1}^2} dx ,
\end{equation}
\begin{equation}
\tilde{\psi}_{-\alpha+1} = \psi_{-\alpha+1} \int \frac{1}{\psi_{-\alpha+1}^2} dx ,
\end{equation}
\begin{equation}
\tilde{\psi}_{\alpha-3} = \psi_{\alpha-3} \int \frac{1}{\psi_{\alpha-3}^2} dx, 
\end{equation}
\begin{equation}
\tilde{\psi}_{\alpha-5} = \psi_{\alpha-5} \int \frac{1}{\psi_{\alpha-5}^2} dx ,
\end{equation}
\begin{equation}
\tilde{\psi}_{\alpha-1} = \psi_{\alpha-1} \int \frac{1}{\psi_{\alpha-1}^2} dx ,
\end{equation}
\begin{equation}
\tilde{\psi}_{\alpha+1} = \psi_{\alpha+1} \int \frac{1}{\psi_{\alpha+1}^2} dx ,
\end{equation}
The action of the ladder operators $B_{III}$ and $B_{III}^{\dagger}$ on the states is given in Figure 6 ( again the action of the lowering operators toward the left and action of the raising operator toward the right ).

\begin{figure}[h]
\begin{equation*}
\xymatrix{
...  & \psi_{-\alpha-1} \ar@/^{5mm}/[r]^{0} \ar@/^{5mm}/[l] &  \psi_{-\alpha+1} \ar@/^{5mm}/[l]^{0} \ar@/^{5mm}/[r]  & ....   &  \psi_{\alpha-5} \ar@/^{5mm}/[l] \ar@/^{5mm}/[r]^{0}   & \psi_{\alpha-3}  \ar@/^{5mm}/[l]^{0} \ar@/^{5mm}/[r]^0   & \psi_{\alpha-1} \ar@/^{5mm}/[l]^{0} \ar@/^{5mm}/[r]^{0}  &   \psi_{\alpha+1}  \ar@/^{5mm}/[l]^{0} \ar@/^{5mm}/[r] &   ...    \\ \\ \\
...  & \tilde{\psi}_{-\alpha-1} \ar@/^{5mm}/[l]  \ar@/^{3mm}/[uuur]   &  \tilde{\psi}_{-\alpha+1}    \ar@/^{3mm}/[uuul] \ar@/^{5mm}/[r]  & ... &  \tilde{\psi}_{\alpha-5} \ar@/^{3mm}/[l]  \ar@/^{3mm}/[uuur]  & \tilde{\psi}_{\alpha-3} \ar@/^{3mm}/[uuul] \ar@/^{5mm}/[uuur] &  \tilde{\psi}_{\alpha-1} \ar@/^{5mm}/[uuul] \ar@/^{3mm}/[uuur]  &  \tilde{\psi}_{\alpha+1}  \ar@/^{5mm}/[uuul] \ar@/^{3mm}/[r]    &  ...
}
\end{equation*}
  \caption{Action of the ladder }  
\end{figure}
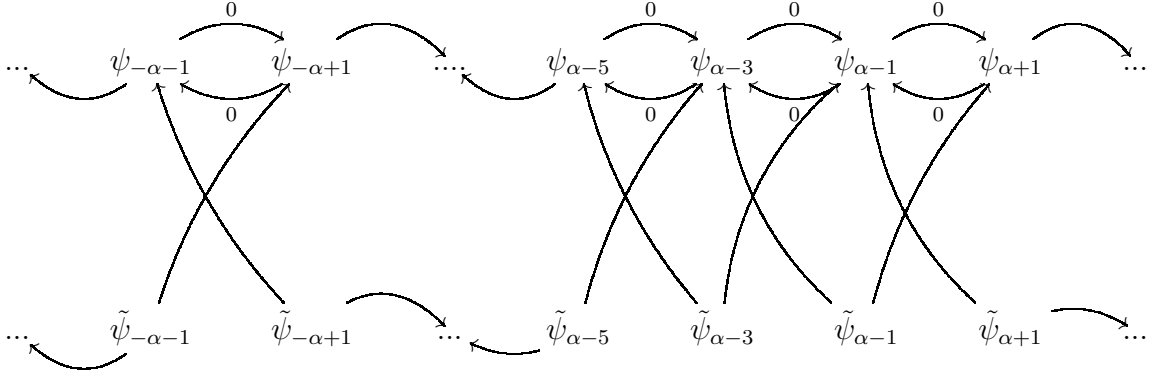

We have the following action for the states of the second chains that relate to the first chain
\begin{equation}
B_{III}^{\dagger} \tilde{\psi}_{\alpha-1} = -4 (\alpha-2) (\alpha-1) \alpha \psi_{-\alpha+1},
\end{equation}
\begin{equation}
    B_{III} \tilde{\psi}_{-\alpha+1} = 4 (\alpha-2)(\alpha-1) \alpha \psi_{-\alpha-1} ,
\end{equation}
\begin{equation}
    B_{III}^{\dagger} \tilde{\psi}_{\alpha-5} = \psi_{\alpha-3},
\end{equation}
\begin{equation}
    B_{III} \tilde{\psi}_{\alpha-3} = - \psi_{\alpha-5},
\end{equation}
\begin{equation}
    B_{III}^{\dagger} \tilde{\psi}_{\alpha-3} = (-2 +2\alpha) \psi_{\alpha-1},
\end{equation}
\begin{equation}
    B_{III}\tilde{\psi}_{\alpha-1} = +1(2-2\alpha) \psi_{\alpha-{\alpha-1}},
 \end{equation}
\begin{equation}
    B_{III}^{\dagger} \tilde{\psi}_{\alpha-1} = \frac{1}{2} \psi_{\alpha+1},
\end{equation}
\begin{equation}
    B_{III} \tilde{\psi}_{\alpha+1} = - \frac{1}{2} \psi_{\alpha-1} .
\end{equation}

There is two type sof states the one that can be used to induced other ones and additional states that need to be added as singlet. The states used to induced other one are $\psi_{-\alpha-1}$, $\psi_{-\alpha+1}$, $\psi_{\alpha-5}$, $\psi_{\alpha+1}$, $\tilde{\psi}_{-\alpha-1}$, $\tilde{\psi}_{-\alpha+1}$, $\tilde{\psi}_{\alpha-5}$ and $\tilde{\psi}_{\alpha+1}$. The state that are added $\tilde{\psi}_{\alpha-3}$, $\psi_{\alpha-1}$, $\tilde{\psi}_{\alpha-3}$ and $\tilde{\psi}_{\alpha-1}$. The different part of the chains are constructed via
\begin{equation}
   \psi_{-\alpha-1-2n} = (B_{III})^n \psi_{-\alpha-1} ,
\end{equation}
\begin{equation}
    \psi_{-\alpha+1+2n}= (B_{III}^{\dagger})^n \psi_{-\alpha+1},
\end{equation}
\begin{equation}
    \psi_{\alpha-5-2n} = (B_{III})^n \psi_{\alpha-5},
\end{equation}
\begin{equation}
    \psi_{\alpha+1+2n} = (B_{III}^{\dagger})^n \psi_{\alpha+1},
\end{equation}
\begin{equation}
   \tilde{\psi}_{-\alpha-1-2n} = (B_{III})^n \tilde{\psi}_{-\alpha-1} ,
\end{equation}
\begin{equation}
    \tilde{\psi}_{-\alpha+1+2n}= (B_{III}^{\dagger})^n \tilde{\psi}_{-\alpha+1},
\end{equation}
\begin{equation}
    \tilde{\psi}_{\alpha-5-2n} = (B_{III})^n \tilde{\psi}_{\alpha-5},
\end{equation}
\begin{equation}
    \tilde{\psi}_{\alpha+1+2n} = (B_{III}^{\dagger})^n \tilde{\psi}_{\alpha+1},
\end{equation}
The action can then be determined for each of those component 
\begin{equation}
    B_{III} \psi_{-\alpha-1-2n}=\psi_{-\alpha-1-2(n+1)},
\end{equation}
\begin{equation}
    B_{III}^{\dagger} \psi_{-\alpha-1-2n}=g_n^{III}(-\alpha-1) \psi_{-\alpha-1-2(n-1)},
\end{equation}
\begin{equation}
    B_{III}^{\dagger} \psi_{-\alpha+1+2n}= \psi_{-\alpha+1+2(n+1)},
\end{equation}
\begin{equation}
     B_{III} \psi_{-\alpha+1+2n} = f_n^{III}(-\alpha+1) \psi_{-\alpha +1+2(n-1)},
\end{equation}
\begin{equation}
    B_{III} \psi_{\alpha-5-2n} = \psi_{\alpha-5-2(n+1)},
\end{equation}
\begin{equation}
    B_{III}^{\dagger} \psi_{\alpha-5-2n} = g_n^{III}(\alpha-5) \psi_{\alpha-5-2(n-1)},
\end{equation}
\begin{equation}
    B_{III}^{\dagger} \psi_{\alpha+1+2n} = \psi_{\alpha+1+2(n+1)},
\end{equation}
\begin{equation}
    B_{III}\psi_{\alpha+1+2n} = f_n^{III}(\alpha+1) \psi_{\alpha+1+2(n-1)},
\end{equation}
and
\begin{equation}
    B_{III} \tilde{\psi}_{-\alpha-1-2n}=\tilde{\psi}_{-\alpha-1-2(n+1)},
\end{equation}
\begin{equation}
    B_{III}^{\dagger} \tilde{\psi}_{-\alpha-1-2n}=g_n^{III}(-\alpha-1) v{\psi}_{-\alpha-1-2(n-1)},
\end{equation}
\begin{equation}
    B_{III}^{\dagger} \tilde{\psi}_{-\alpha+1+2n}= \tilde{\psi}_{-\alpha+1+2(n+1)},
\end{equation}
\begin{equation}
     B_{III} \tilde{\psi}_{-\alpha+1+2n} = f_n^{III}(-\alpha+1) \tilde{\psi}_{-\alpha +1+2(n-1)},
\end{equation}
\begin{equation}
    B_{III} \tilde{\psi}_{\alpha-5-2n} = \tilde{\psi}_{\alpha-5-2(n+1)},
\end{equation}
\begin{equation}
    B_{III}^{\dagger} \tilde{\psi}_{\alpha-5-2n} = g_n^{III}(\alpha-5) \tilde{\psi}_{\alpha-5-2(n-1)},
\end{equation}
\begin{equation}
    B_{III}^{\dagger} \tilde{\psi}_{\alpha+1+2n} = \tilde{\psi}_{\alpha+1+2(n+1)},
\end{equation}
\begin{equation}
    B_{III} \tilde{\psi}_{\alpha+1+2n} = f_n^{III}(\alpha+1) \tilde{\psi}_{\alpha+1+2(n-1)},
\end{equation}
This complete the description of the 2-chains representations for the Laguerre EOPs of type III.

\par
%
%
\section{Conclusion}

In this paper, we presented a new approach for the construction of general solutions of the differential equation associated with exceptional orthogonal polynomials of Laguerre  $X_1$ for the type I, II and III. This allows to obtain induced type representations that take the form of two chains interconnected via the action of the cubic Heisenberg-Weyl algebra. This provides an interesting connection between between functions and representation theory of polynomial algebras. The representations are infinite dimensional and contain the corresponding EOPs $X_1$ of type I, II and III. The results provides also Rodrigues type formula that are of interest.

We have presented explicit formula for the representations and action of the generators and as well presented graphical descriptions of the 2-chains. 

The paper also develop novel aspects of polynomial Heisenberg algebras and the construction of several identities for monomials of the generators which are also of interest for further applications of those algebraic structures. 

\par
%
%
.\par
%
%
\section*{Acknowledgments}

IM was supported by the Australian Research Council Fellowship FT180100099. I.M. thanks the University of Warsaw, the Centre de Recherches Math\'ematiques and Universit\'e de Montr\'eal for their hospitality. He thanks also Prof. Michel Grundland for his comments on the paper. \par
%
%

\par
%
%

\end{document}